\documentclass[
reprint,
superscriptaddress,
nofootinbib,
nobibnotes,
 amsmath,amssymb,
 aps,
prc,
]{revtex4-1}

\usepackage{graphicx}% Include figure files
\usepackage{dcolumn}% Align table columns on decimal point
\usepackage{bm}% bold math
\usepackage{hyperref}% add hypertext capabilities
\newcommand{\E}{E_{\nu}}
\newcommand{\Erec}{\overline{E}_{\nu}}
\newcommand{\ddCS}{\frac{\,d\sigma}{\,d\omega\,d\cos\theta}}
\newcommand{\ddCSEdep}{\frac{\,d\sigma\left(\E\right)}{\,d\omega\,d\cos\theta}}
\def\etal.{et\penalty50\ al.}
\begin{document}

\preprint{APS/123-QED}

\title{Mean field approach to reconstructed neutrino energy distributions in accelerator-based experiments}% Force line breaks with \\

\author{A. Nikolakopoulos}
\email{alexis.nikolakopoulos@ugent.be}
\affiliation{Department of Physics and Astronomy, Ghent University, Proeftuinstraat 86, B-9000 Gent, Belgium}
\author{M. Martini}
\affiliation{ESNT, CEA, IRFU, Service de Physique Nucléaire, Universit\'e Paris-Saclay, 91191 Gif-sur-Yvette, France}
\author{M. Ericson}
\affiliation{Universit\'e de Lyon, Univ. Lyon 1, CNRS/IN2P3, IPN Lyon,F-69622 Villeurbanne Cedex, France}
\affiliation{Physics Department, Theory Unit, CERN, CH-1211 Geneva, Switzerland}
\author{N. Van Dessel}
\affiliation{Department of Physics and Astronomy, Ghent University, Proeftuinstraat 86, B-9000 Gent, Belgium}
\author{R. Gonz\'{a}lez-Jim\'{e}nez}
\affiliation{Grupo de F\'isica Nuclear, Departamento de Estructura de la Materia, F\'isica T\'ermica y Electr\'onica, \\
 Universidad Complutense de Madrid, CEI Moncloa, 28040  Madrid, Spain}
\affiliation{Department of Physics and Astronomy, Ghent University, Proeftuinstraat 86, B-9000 Gent, Belgium}
\author{N. Jachowicz}
\email{natalie.jachowicz@ugent.be}
\affiliation{Department of Physics and Astronomy, Ghent University, Proeftuinstraat 86, B-9000 Gent, Belgium}

\date{\today}% It is always \today, today,
             %  but any date may be explicitly specified june 19

\begin{abstract}
\begin{description}
\item[Background]  
The reconstruction of the neutrino energy is crucial in oscillation experiments that use interactions with nuclei to detect the neutrino.
 The common reconstruction procedure  is based on the kinematics of the final-state lepton. 
The interpretation of the reconstructed energy in terms of the real neutrino energy must rely on a model for the neutrino-nucleus interaction. 
The Relativistic Fermi Gas (RFG) model is frequently used  in these analyses. 
%has shown to be difficult to interpret.
\item[Purpose] We examine  the effects of nuclear structure and dynamics going beyond a Fermi gas model on the reconstruction procedure.
%, for which it has been shown that the reconstructed energy is statistically a rather good approximation to the actual energy.
\item[Method] In the Hartree-Fock (HF) model for quasielastic nucleon knockout, the bound nucleon wave functions are obtained through a calculation using an effective nucleon-nucleon force.
The final-state wave function is constructed from continuum states in the same potential which have the correct asymptotic behavior.
The Continuum Random Phase Approximation (CRPA) model extends the HF approach taking long range correlations into account in a self-consistent way.  
\item[Results]
Considering only single-nucleon processes, the distributions of (reconstructed) neutrino energies obtained within the HF-CRPA approach are compared with the results of the RFG, a relativistic plane wave impulse approximation (RPWIA) calculation, and the RPA+np-nh model of Martini~\etal.
\item[Conclusions]
 We find that the distributions of reconstructed energies for a fixed incoming energy in the HF-CRPA display additional strength in the lower reconstructed energy tails compared to models without elastic distortion of the outgoing nucleon and the mean field description of the initial nucleon.
This asymmetry redistributes strength from higher to lower values of the reconstructed energy.
The mean field description of the nuclear dynamics results in a reshaping of the reconstructed energy distribution that cannot be accounted for in a plane wave impulse approximation model, even by modifying ad hoc parameters such as the binding energy.
In particular it is shown that in the RFG calculations there is no value of the binding energy which is able to reproduce the entire T2K $\nu_\mu$ oscillated spectrum as calculated in HF-CRPA. 
\end{description}
\end{abstract}

\maketitle

%\tableofcontents

\section{\label{sec:introduction}Introduction}

The main goal of  accelerator-based neutrino experiments  is the determination of the neutrino oscillation parameters.
 The oscillation probability depends on the ratio of the distance traveled by the neutrino to its energy, therefore the determination of the neutrino energy distribution in a detector is crucial in the oscillation analysis.
Due to the broad energy distribution of the neutrino beams, determining the energy of the neutrinos at a detector is a non-trivial task, which depends on the model used to describe the neutrino interaction with the target nucleus.
In a detector one observes the charged-current scattering of a neutrino off a nucleus where a single final-state lepton is detected. The hadronic final state is in many cases not, or not fully, accessible. 
These events are dubbed quasielastic (QE)\emph{-like}. This means that CCQE scattering is considered the dominant reaction mechanism, but depending on the kinematics other mechanisms should be taken into account.
 The experimentally measurable $E_l$ and $\cos\theta$, the energy and the scattering angle of the final lepton, are then used to define the reconstructed energy in the QE(-like) interaction \cite{MB:QECS,T2k:rec}
\begin{equation}
\label{eq:erecbind}
\overline{E}_{\nu}= \frac{2M^{\prime}_nE_{l}-({M^{\prime}_n}^2 
+m_l^2 -M_p^2)}{2(M^{\prime}_n - E_l +P_l\cos\theta)},
\end{equation}
%\begin{equation}
%\label{eq:erecbind}
%\overline{E}_{\nu}= \frac{2M^{\prime}_nE_{l}-(M^{\prime}_n^2 +m_l^2 -M_p^2)}{2(M^{\prime}_n - E_l +P_l\cos\theta)},
%\end{equation}
%\begin{equation}
%\label{eq:erecbind}
%\overline{E}_{\nu}= \frac{2M^{\prime}_nE_{l}-(M^{\prime}^2_n +m_l^2 -M_p^2)}{2(M^{\prime}_n - E_l +P_l\cos\theta)},
%\end{equation}
where $M_n^{\prime}=M_n-E_B$ is the adjusted neutron mass, with $M_n$($M_p$) the neutron(proton) rest mass, and $E_B$ a chosen binding energy.
The lepton rest mass and momentum are $m_l$ and $P_l$ respectively.\\
 
The reconstructed energy $\overline{E}_{\nu}$ defined in this way is the energy of the incoming neutrino in a genuine QE charged-current scattering off a neutron at rest, adjusted for binding through the parameter $E_B$. Notice that there is a certain ambiguity in the definition of binding energy of the nucleon.
 Different definitions of binding energy, or similar quantities such as missing energy or single particle energy, require some caution in their use. A thorough discussion of different definitions and interpretations of binding energy in leptonic interactions with the nucleus is given in Ref.~\cite{Bodek:Binding}.\\

The reconstructed energy $\overline{E}_{\nu}$ as defined above is easy to calculate but it is not a good estimate of the true neutrino energy on an event to event basis, as it considers the dynamics of the nucleons in a nucleus only through the quantity $E_B$.

After binning the data in terms of the reconstructed energy, the true energy distribution has to be recovered.
 This requires a theoretical model for the interaction which can provide the conditional probability of an event having true energy $E_\nu$, given the reconstructed energy $\overline{E}_\nu$.
 Non-inclusion of a realistic cross section model, or possible reaction mechanisms, can introduce a strong bias in the oscillation parameters.
 
 Indeed in Refs.~\cite{MartiniModel:2009, MartiniModel:2010}, Martini~\etal.\  pointed out the importance of additional strength in the data due to multinucleon excitations which in a \^{C}erenkov detector are indistinguishable from pure quasielastic reactions.
The influence of multi-nucleon emission on the reconstructed energy  was investigated in Refs.~\cite{Martini2012,GiBUU:2p2h,Nieves:Unfolding}.
Further efforts on studying the reconstruction procedure have mainly focused on interactions other than QE one nucleon knockout.
Reactions that contribute to \emph{'QE-like'} events, such as multi-nucleon knock-out, misidentification or reabsorption of pions, and excitation of resonances which lead to the same experimental signal were shown to significantly affect the reconstructed energy distributions \cite{LeitnerandMosel,Lalakulich:QErec}.
The effect of 2p-2h events on the oscillation analysis and the electron-like event excess found by MiniBooNE for low reconstructed energies \cite{MBexcess:2009} (and recently confirmed \cite{MBexcess:2018}) has also been examined \cite{Martini:recooscill,Ericson:Reco}.
Ankowski~\etal.~\cite{Ankowski:BetterE} on the other hand have analyzed the reconstruction problem in a spectral function formalism in which they include the distortion of the emitted nucleon in an effective way.

Restricting to the single-nucleon knock-out mechanism, we present a comparison between the outcomes of different models. These models include the mean field continuum random phase approximation  model  (HF-CRPA) developed by the Ghent group, which has  been  successful in reproducing the inclusive electron scattering data, including in regions of low energy transfer \cite{PRC92, CRPAmod}.
 We compare the results of this approach to those obtained with the relativistic Fermi gas (RFG) which is commonly used in the experimental analyses.
We also show the result of the relativistic plane wave impulse approximation (RPWIA), in which the bound nucleons are described by relativistic mean field (RMF) wave functions \cite{WaleckaRMF}, the outgoing nucleon however is described by a plane wave. 
We will show that the RFG, although successful in reproducing integrated quantities such as the total cross section, is unable to explain the distribution of reconstructed energies predicted by the mean field approach.
Distortion of the final state nucleon, which is neglected in RFG and RPWIA models, is seen to affect the reconstructed energy distribution considerably.
 
This paper is structured as follows:
In section \ref{sec:level1}, we first compare the distribution of neutrino energies in the CRPA with the results of the RFG and RPWIA models for fixed lepton observables, i.e.~for  fixed values of the reconstructed energy. We then introduce the distribution of reconstructed energies for a given incoming energy.
 Section \ref{sec:level2} in which the results are presented  is divided into three subsections.
We compare the conditional probability distributions of reconstructed energies given a certain real neutrino energy in subsection \ref{sec:level21}.
Secondly, in subsection \ref{sec:level23} we compare the distribution of real and reconstructed energies given the MiniBooNE flux.
Then in subsection \ref{sec:level24}, we compare the predicted distributions of reconstructed energies for the T2K near and far detector fluxes using a simple two-neutrino oscillation probability. 
The conclusions are presented in section \ref{sec:conclusions}.

\section{\label{sec:level1}Distributions of reconstructed energies}
In this section, we present the method used to determine the number of events for a certain reconstructed energy, given the cross section and the predicted flux. 
We will closely follow the approach of Refs. \cite{Martini2012,Martini:recooscill}, an equivalent discussion is given in Refs. \cite{Lalakulich:QErec,Nieves:Unfolding}.
In the first subsection the flux-weighted cross section for fixed lepton observables, i.e., for fixed reconstructed energy, denoted as $f\left(\E,E_l,\cos\theta\right)$, is examined. 
From $f\left(\E,E_l,\cos\theta\right)$ one can appreciate the range of energy values $E_{\nu}$ that will be reconstructed to the same $\Erec$ given a certain $\cos\theta$ and $E_l$. 
Then in \ref{subsecEErec} the single-differential cross section $\frac{d\sigma(\E,\Erec)}{d\Erec}$,
is derived. This quantity gives direct access to the number of events in bins of $\Erec$, which is what is reported by experiments. 

We compare a number of different models in this paper.
Firstly the mean-field Continuum Random Phase Approximation model, to which we will refer as HF-CRPA~\cite{CRPAmod,PRC92}. In this model the nucleus is described in a mean field generated by a self-consistent Hartree-Fock method, using the extended Skyrme force (SkE2) \cite{Skyrme,WAROQUIER1983} for the nucleon-nucleon interactions.
The CRPA takes long-range correlations into account consistently by using the same nucleon-nucleon interaction \cite{RYCKEBUSCH1988,RYCKEBUSCH1989}.
The knocked-out nucleon is modeled as continuum state of the residual nucleus.
By using the same HF potential for the calculation of the outgoing nucleon wave function, the initial and final states are orthogonal.
The model has been tested  for electron and neutrino scattering data in \cite{RYCKEBUSCH1989,CRPAmod,Pandey:2016}. A comparison with the results of the model of Martini~\etal.\ was presented in Ref.~\cite{Martini:Jachowicz}. 
We summarize here some of the general differences of the CRPA approach compared to the RPA model of  Martini~\etal.~\cite{MartiniModel:2009, MartiniModel:2010}.
When the HF model is compared to the bare (i.e.~no RPA) approach of Martini~\etal.\ it is found that the double differential cross section in the HF model is quenched, featuring more strength for larger energy transfers.
The RPA model of Martini leads to a  quenching of the cross section, for a large part due to the admixture of nucleon states with the delta resonance, which makes the double differential cross section in the RPA comparable to the results found in the HF model in the QE-region.
The effect of the CRPA is twofold, firstly there is a slight suppression of the cross section in the QE-region up to the QE-peak. On the other hand, for low energy transfers, the CRPA approach leads to an enhancement of the cross section from the contributions of giant resonances~\cite{Martini:Jachowicz}, while these collective states are not included in the RPA approach of Ref.~\cite{martiniEricson:2014}.

Secondly, we consider the RFG, described in the appendix of \cite{Amaro:2005}, which includes Pauli-blocking \cite{PauliBlocking}.
In the RFG we include a shift by substitution of $\omega\rightarrow\omega_{eff} = \omega - E_s$, to account for the displacement of the quasielastic peak.

The third model is the RPWIA model, in which the bound state nucleons are described by RMF wave functions \cite{WaleckaRMF}, however the outgoing nucleon is modeled by a plane wave. The cross section is then obtained as an incoherent sum over the different nucleon shells which have a specific value of the missing energy~\cite{Bodek:Binding}, thereby unambiguously determining the four momentum of the outgoing nucleon. 

Finally, we also show results of the model with relativistic corrections of Martini~\etal.~\cite{MartiniModel:2009,MartiniModel:2010,MM2011}, referred to as RPA, and RPA+np-nh for the model including multi-nucleon excitations. In this model the nuclear initial state is described using a local Fermi gas.

\subsection{Fixed lepton observables}
When considering an interaction where the final state lepton has scattering angle $\cos\theta$ and energy $E_l$, and using $\omega = \E-E_l$, the probability that the incoming neutrino had energy $\E$ is proportional to
\begin{equation}
\label{eq:fdist}
f(\E,E_l,\cos\theta)=C\Phi(\E)\left.\ddCSEdep\right\rvert_{\omega=\E-E_l}.
\end{equation}
Here $C$ is a normalization constant, chosen to be the inverse of the total flux, $\left(\int\,d\E\Phi(\E)\right)^{-1}$.
As the reconstructed energy is completely determined by $\cos\theta$ and $E_l$, $f\left(\E,E_l,\cos\theta\right)$ is proportional to the probability of a neutrino having energy $\E$ for certain fixed $\Erec\left(\cos\theta,E_l\right)$ as determined in Eq. (\ref{eq:erecbind}).

\begin{figure}[tb]
\begin{center}
\includegraphics[width=0.48\textwidth]{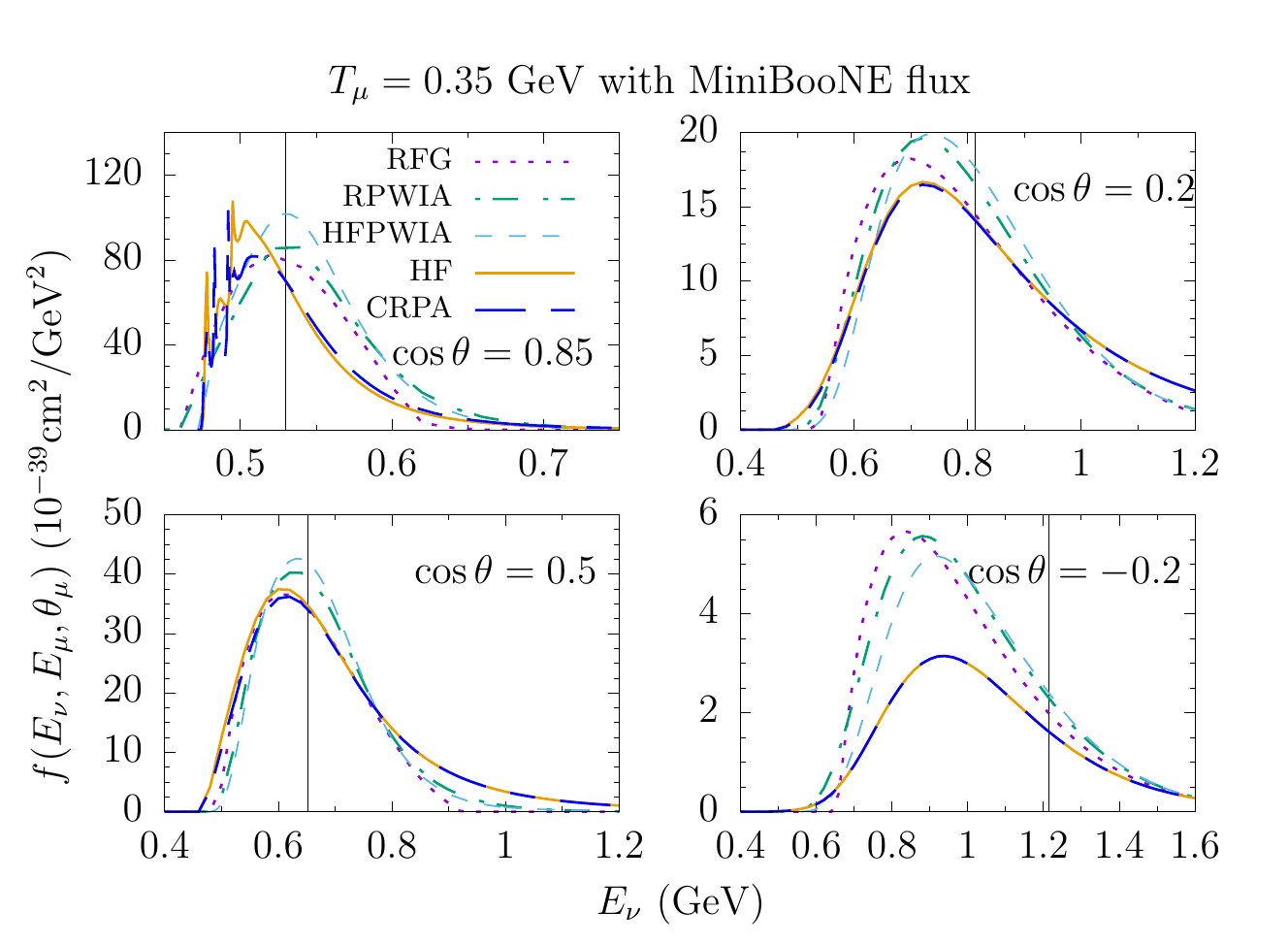}
\caption{Comparison of the double differential cross section for fixed lepton observables, weighted by the MiniBooNE flux \cite{MBflux:2009}, $f(\E,E_{\mu},\cos\theta_{\mu})$ defined in Eq. (\ref{eq:fdist}).
We compare the results of the HF-CRPA model with the RFG description, the relativistic plane wave impulse approximation (RPWIA), and the Hartree-Fock plane wave impulse approximation (HFPWIA). The vertical lines denote the reconstructed energy for the selected kinematics $\Erec(T_\mu,\cos\theta)$ defined in Eq.~(\ref{eq:erecbind}) with $E_B = 34~\mathrm{MeV}$.}
\label{fig:tripdens_RFG_35}
\end{center}
\end{figure}

In Fig. \ref{fig:tripdens_RFG_35}, we compare the HF-CRPA results to those obtained with the RFG, RPWIA, and HFPWIA.
The HFPWIA is identical to the full HF approach, except that a plane wave is used for the outgoing nucleon wave function.
In the RFG we use a shift in the energy transfer of $34~\textrm{MeV}$ by substituting $\omega \rightarrow \omega - E_s$.

The initial momentum distribution of the bound nucleons is different in these approaches. 
In the RFG model all momentum states up to the Fermi momentum are occupied. 
The Pauli-blocking correction leads to a reduction of the cross section at low energy transfer, by using a hard cut off when the outgoing nucleon has a momentum lower than the Fermi momentum.
In the mean field approaches the momentum distribution is determined by the wave functions of the bound nucleons.
It smoothly goes to zero allowing higher values of missing momentum to contribute to the cross section.
When looking at the double differential cross sections for fixed incoming energy and lepton scattering angle, this leads to the RFG cross section sharply going to zero.
On the contrary in the mean field models there is a larger region in which the nuclear response is non zero, reflected in a high-$\omega$ tail for the double differential cross section.

It is well known, through comparison to inclusive electron scattering data, that the RPWIA model is significantly improved by incorporating distortion of the outgoing nucleon wave function.
By using a solution of the Dirac equation in the same mean field for the outgoing nucleon wave function, the height of the QE peak is reduced,
 in favor of the high-$\omega$ tail \cite{Scaling_RFSI,SuSav2}.

These considerations allow us to explain the differences between the models, we start our discussion by considering the scattering angle $\cos\theta=0.5$ in Fig.~\ref{fig:tripdens_RFG_35}.
It is known that the RFG model gives a good description of the QE peak in inclusive electron scattering in this kinematic region when the Fermi momentum and binding energy are tuned accordingly \cite{Moniz}.
The HF-CRPA approach and RFG agree in their description of the peak position and strength, but the HF-CRPA models feature a more prominent high-energy tail.
This tail originates from the high-$\omega$ tail in the double differential cross section.
The PWIA models overestimate the peak as predicted by the RFG, and underestimate the tail in the HF-CRPA description. 
The distortion of the outgoing nucleon wave function is needed to shift strength from the peak to the tail as in the full HF-CRPA description.

For $\cos\theta = 0.85$, we see that the CRPA leads to a slight reduction of the HF cross section along the peak, accompanied by an increase in the tail. 
This forward scattering region, for low energy transfers, is where the more realistic treatment of nuclear dynamics in the HF-CRPA approach leads to a different shape of the cross section.
This region is also where the cross section is largest, and will determine the peak position the distribution of reconstructed energies, which will be shown in the following section.
The reconstructed energy tends to match the peak of the RFG and PWIA cross sections, but describes the average of the HF-CRPA distributions instead of the peak.
The distortion of the outgoing nucleon wave function in the same mean field potentials used for the bound state makes sure that the final and initial states are orthogonal.
This treatment leads to a strong reduction of the cross section and a characteristic shape which compares favorably to inclusive electron scattering data \cite{RYCKEBUSCH1989,CRPAmod,Pandey:2016}.
 
For more backward scattering processes, the height of the peak in the HF-CRPA calculation is reduced.
The mean field tail is less prominent, and the PWIA and RFG cross sections are comparable up to a shift for larger energies, seen for $\cos\theta = -0.2$.
The strong reduction of the cross section in HF-CRPA for more backward angles comes from the distortion of the outgoing wave function.
This can be seen by comparing the PWIA results that lack this distortion and are comparable in magnitude.
We have also confirmed this through direct comparison of the HF-CRPA to the full RMF model of Refs.~\cite{RMF_CCQE, Scaling_RFSI}. 

The corresponding results for the same kinematic conditions as in Fig.~\ref{fig:tripdens_RFG_35} obtained with the RPA and RPA+np-nh models can be found in Fig.~3 of Ref.~\cite{Martini2012}.

\subsection{The single differential cross section $\frac{\,d\sigma(\E,\Erec)}{\,d\Erec}$}\label{subsecEErec}
In this section, we introduce the single differential cross section $\frac{\,d\sigma(\E,\Erec)}{\,d\Erec}$ \cite{Nieves:Unfolding}.
This quantity is denoted  $d(\E,\Erec)$ as in Ref.~\cite{Martini:recooscill}, and is proportional to the probability that a reaction giving rise to a reconstructed energy $\Erec$ was induced by a neutrino with energy $\E$.

The number of events with reconstructed energy $\Erec$, for a given neutrino flux $\Phi\left(\E\right)$, can then be obtained from $d\left(\E,\Erec\right)$ as
\begin{equation}
\label{eq:fluxd}
N(\Erec) = \int\,d\E\Phi(\E)d(\E,\Erec).
\end{equation}
It is essentially this quantity that is measured by experiments after reconstruction from the outgoing lepton variables.

Note also that the number of events in terms of real energies is given by 
\begin{equation}
\label{eq:CStot}
N(\E) = \Phi(\E)\int\,d\Erec d(\E,\Erec).
\end{equation}
This is shown in \cite{Nieves:Unfolding}, and should be clear from the construction of $d(\E,\Erec)$.
In this paper we always use the energy normalized flux $\Phi(\E)\left(\int\,d\E\Phi(\E)\right)^{-1}$ when computing $N(\Erec)$ and $N(\E)$ such that they have the unit $\mathrm{cm}^2~\mathrm{GeV}^{-1}$.
\subsubsection{Construction of $d(\E,\Erec)$}
The number of events for certain  kinematics $(E_{\nu}, \omega, \cos\theta)$ can be written straightforwardly as the product of the flux $\Phi(E_{\nu})$ with the double-differential cross section.
%\begin{align}
% &p(E_{\nu},\omega,\cos\theta)\,d E_{\nu}\,d\omega\,d\cos\theta = \nonumber \\
%&\ddCS\left(\E,\omega,\cos\theta\right)\Phi(\E)\,d E_{\nu}\,d\omega\,d\cos\theta.
%\end{align}
In order to construct $d(\E,\Erec)$, proportional to the number of events with reconstructed energy $\Erec$, the cross section is transformed to the variables $(\E, E_l, \Erec)$.
The lepton energy is determined by $\omega = \E - E_l$, and the scattering angle $\cos\theta(\E,E_l)$ is obtained as a function of $E_l$ and $\Erec$ by inverting Eq.~(\ref{eq:erecbind}).
The transformation is then given by
\begin{equation}
\label{eq:DDCS_Erec}
\frac{\,d\sigma}{\,dE_l\,d\Erec} = \mathcal{J}\left.\ddCS\right\vert_{\omega=\E-E_l,\cos\theta(E_l,\Erec)},
\end{equation}
with $\mathcal{J}$  the Jacobian determinant.
%One can assume a general case for some reconstruction procedure $\Erec=g(E_l,\cos\theta)$, i.e. he reconstructed energy depends on the measurable kinematic variables.
When the lepton energy is determined by
$\omega = \E - E_l$,
the only dependence on the reconstructed energy enters through $\cos\theta(E_l,\Erec)$.
One then finds that the Jacobian is given by 
\begin{equation}
\left\rvert\frac{\partial \cos\theta}{\partial \Erec} \right\rvert= \frac{M^\prime_n E_l-((M^{\prime}_n)^2+m_l^2-M_p^2)/2}{\Erec^2P_l}.
\end{equation}
When equating the neutron and proton mass, and omitting $E_B$, i.e. $M^{\prime}_n=M_p=M$, one obtains  
 the same expression used in Refs. \cite{Martini:recooscill, Nieves:Unfolding}.

\begin{figure}[tb]
\begin{center}
\includegraphics[width=0.48\textwidth]{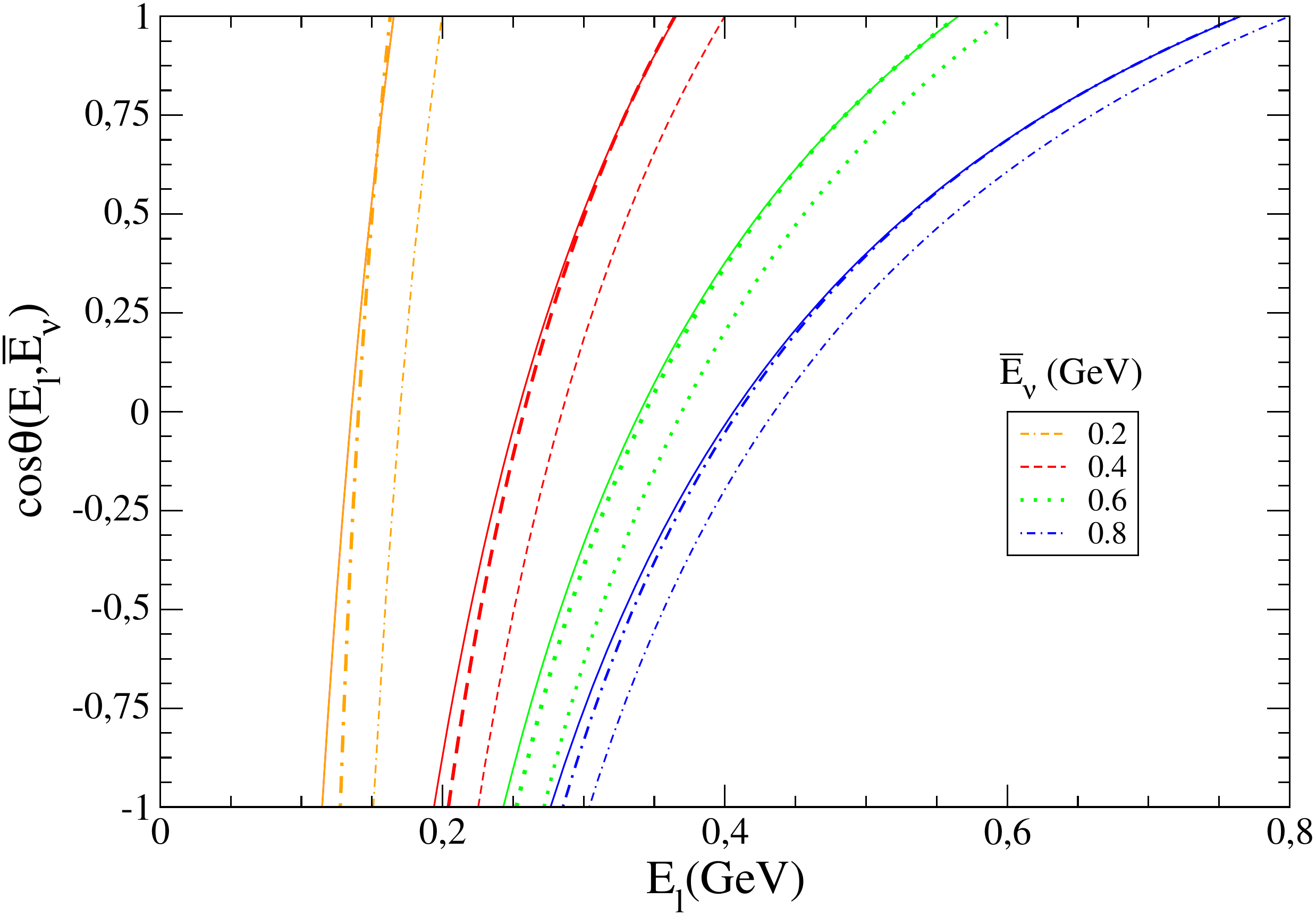}
\caption[The effect of the separation energy and rest mass of the final state lepton on $\cos\theta(E_l,\Erec)$.]{The effect of the binding energy and rest mass of the final state lepton on $\cos\theta(E_l,\Erec)$. The dashed lines are for a final state muon, where the thin dashed lines do not include a binding energy and the thick dashed lines are for $E_B=34~\textrm{MeV}$. The solid lines are the solutions for a final state electron using the same binding energy.}
\label{fig:cos}
\end{center}
\end{figure}

The distribution $d(\E,\Erec)$ is now defined by integration over the lepton energy
\begin{align}
\label{eq:ddef}
& d(\E,\Erec) \nonumber \\ 
& = \int_{E_l^-}^{E_l^+} \,d E_l \mathcal{J}\left.\ddCS\right\vert_{\omega=\E-E_l,\cos\theta(E_l,\Erec)}.
\end{align}
The integration bounds are given by the most extreme values of the lepton energy that reconstruct to a certain $\Erec$, i.e.~the energies for which $\cos\theta(E_l,\Erec) = \pm 1$.

In Fig.~\ref{fig:cos}, the function $\cos\theta(E_l,\Erec)$ is plotted for different values of $\Erec$.
The dashed lines represent the solution obtained with the reconstruction formula when $E_B=0$.
The thick dashed lines are obtained with $E_B=34~\textrm{MeV}$.
Both dashed lines are solutions for a final state muon.
The solution for electrons, also with $E_B=34~\textrm{MeV}$, is shown by the solid line.
This figure should be interpreted in the light of Eq.~(\ref{eq:ddef}).
The integration range can be determined from the $E_l$-axis, the cross section of Eq.~(\ref{eq:DDCS_Erec}) is integrated for values of $E_l$ between $E^+_l$ and $E^-_l$, given by the points for which $\cos\theta(E_l,\Erec)=\pm1$.
For each $E_l$ in this range, the angle by which the integrand is considered is given by $\cos\theta(E_l,\Erec)$.
For a certain $\Erec$ the line thus corresponds to the total leptonic phase-space that can be reconstructed to this value.  
Integration of the cross section for a fixed energy $\E$ along this line yields $d(\E,\Erec)$, and the total cross section is recovered as in Eq. (\ref{eq:CStot}), which can be understood from Fig. \ref{fig:cos} as the available leptonic phase-space being covered by lines of different $\Erec$ values.

\section{Results\label{sec:level2}}
As seen in the previous section, where the flux-weighted distributions of energies $f(\E,E_l,\cos\theta)$ defined in Eq.~(\ref{eq:fdist}) were presented, the mean field approaches feature more strength in the tails of the cross section.
We first show the effect this has on the distribution $d(\E,\Erec)$.
Then we will illustrate how the experimentally measurable distributions of reconstructed energies differ.

\subsection{The single differential cross section $d\left(\E,\Erec\right)$\label{sec:level21}}

\begin{figure}[tb]
\begin{center}
\includegraphics[width=0.48\textwidth]{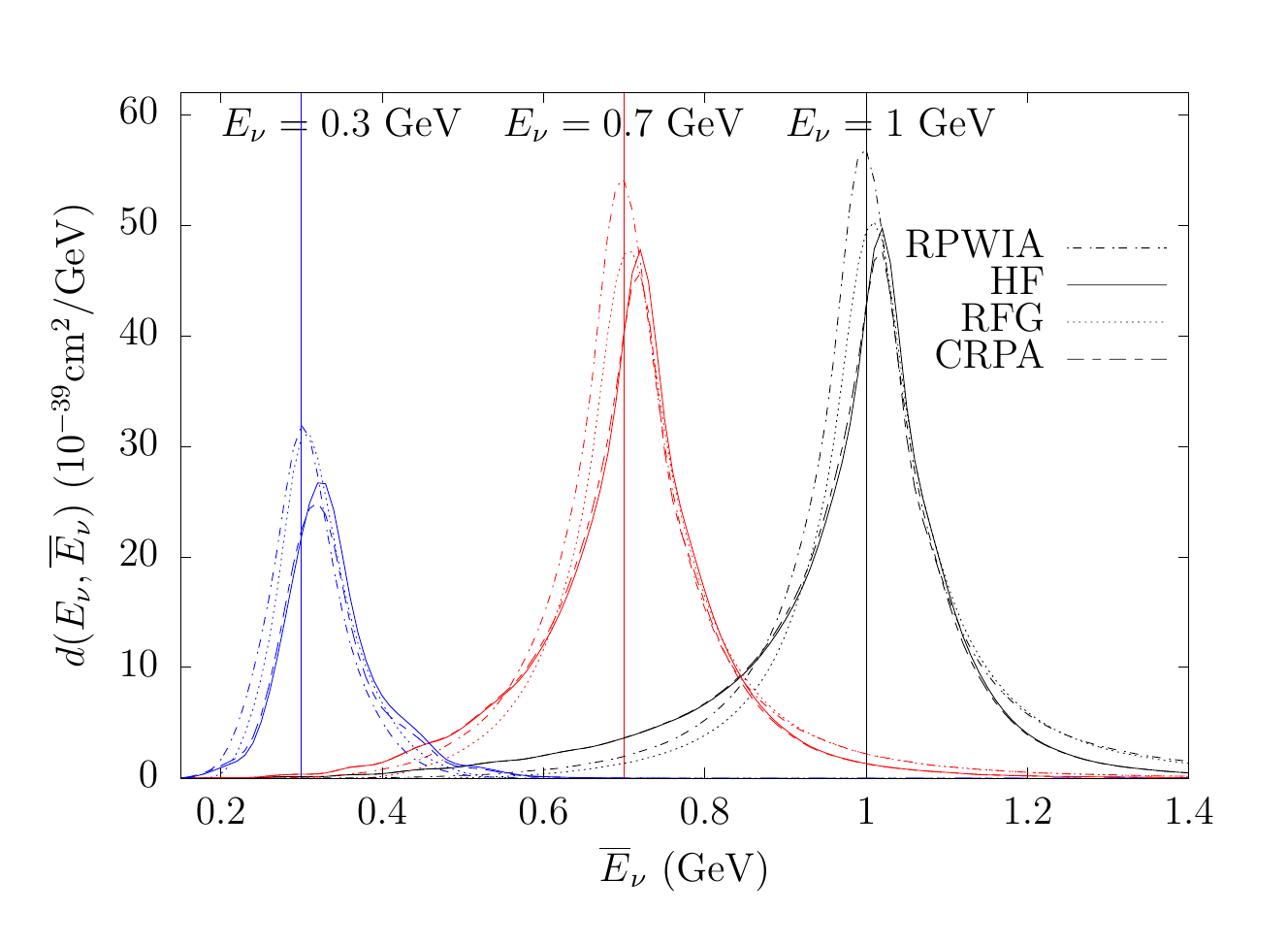}
\caption{Comparison of $d(\E,\Erec)$ obtained with the CRPA, HF, RPWIA and RFG models. The reconstructed energy is defined by Eq. (\ref{eq:erecbind}) with $E_B = 34~\mathrm{MeV}.$
The RFG model includes an energy shift of $E_s = 34~\mathrm{MeV}$}
\label{fig:RPWIA_RFG_HF}
\end{center}
\end{figure}
We compare $d\left(\E,\Erec\right)$ obtained in different approaches. 
The comparison of the HF-CRPA with the RFG and RPWIA models is shown in Fig.~\ref{fig:RPWIA_RFG_HF}.
We use $\Erec$ as defined in Eq.~(\ref{eq:erecbind}), with $E_B = 34~\mathrm{MeV}$ as used in MiniBooNE \cite{MB:QECS}.
The RFG cross section includes a shift with $E_s = 34~\mathrm{MeV}$ chosen to match $E_B$.

We see that the CRPA leads to a slight decrease of the peak strength compared to the HF that is most apparent at lower neutrino energies.
In any case, the general shape of $d(\E,\Erec)$ is similar in the HF and CRPA approaches, which is why we will refer to both approaches simultaneously as HF-CRPA.

The peaks of the HF-CRPA models are slightly displaced to larger reconstructed energies, while the peaks in the RFG and RPWIA are centered around the reconstructed energy.
The peak position for the RPWIA and RFG models is determined by the average binding energy of the nucleon.
The peak position in the RFG reproduces the reconstructed energy because we choose $E_s = E_B$.
In the RPWIA the effect of binding is determined by the separation energy of the two discrete shell-model states, this quantity is not directly comparable to the energy shift used in the RFG.
The binding in the HF-CRPA approach is in the first instance also determined by the single-particle energies of the bound neutrons.
 One can appreciate that for sufficiently large energies ($\E = 0.7~\mathrm{GeV}$ in Fig.~\ref{fig:RPWIA_RFG_HF}) the value for which $\E = \Erec$ tends to match the \emph{average} of $d(\E,\Erec)$. 
However the peak is displaced further to the right because of the more accurate treatment of the low-$\omega$ cross section and the distortion of the outgoing nucleon. 
Indeed as can be appreciated from the discussion in Ref. \cite{Martini:Jachowicz}, the HF-CRPA approach leads to a broader cross section, in which the peak of the cross section is quenched, an effect that goes hand in hand with a more prominent high-$\omega$ tail. 
The peak of $d(\E,\Erec)$ is mainly determined by the forward scattering cross section.
Looking at the results for $\cos\theta = 0.85$ in Fig.~\ref{fig:tripdens_RFG_35} one sees that the more realistic shape of the HF-CRPA cross section indeed leads to a peak at energies smaller than $\Erec$, while the RPWIA and RFG approaches have a peak along the reconstructed energy.

\begin{figure}[tb]
\centering
\includegraphics[width=0.48\textwidth]{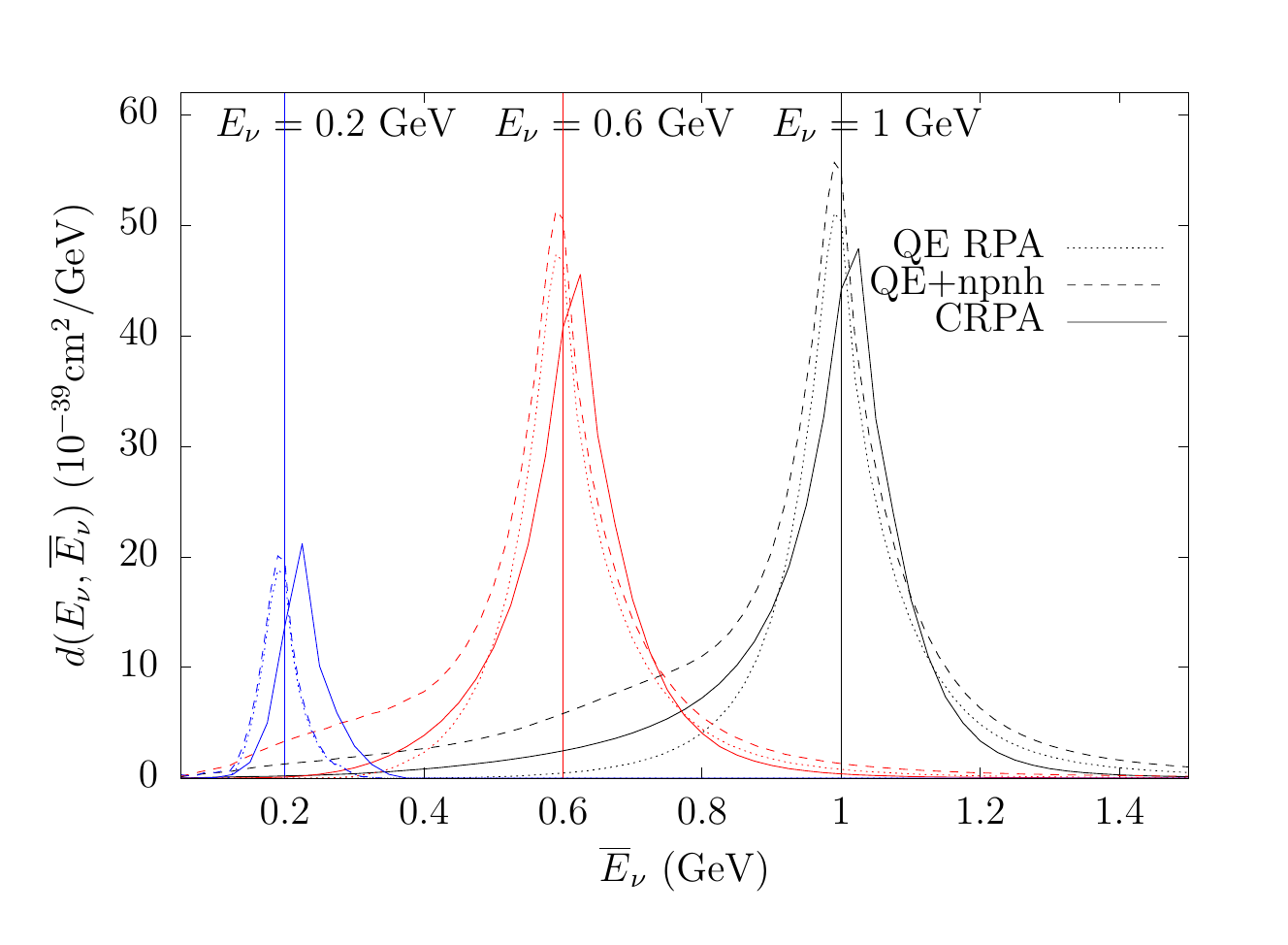}
\caption{Comparison of $d(\E,\Erec)$ in the CRPA (solid lines) approach compared to the results obtained by Martini~\etal.~\cite{Martini:recooscill} in the RPA approach (dotted lines) and RPA+np-nh (dashed lines).
The distributions are calculated for electron neutrinos, the results of Martini are obtained with $E_B=0~\mathrm{MeV}$, while the CRPA results do include a binding energy of $E_B=34~\mathrm{MeV}$ in the reconstructed energy.}
\label{fig:MMvsCRPA}
\end{figure}

The RFG model lacks the low-$\Erec$ tails present for sufficiently high energies in the mean-field approaches.
These originate from the broader momentum distributions in the mean field models, which lead to a larger phase space in which the nuclear response is non-zero.
The HF-CRPA models however feature far more strength in the low-$\Erec$ tail than the RPWIA.
The general features of $d(\E,\Erec)$ in the HF-CRPA model, namely the shift of the peak, large low-$\Erec$ tail and quenched tail to the right of the peak, can be understood by considering that the distortion of the outgoing nucleon reduces and shifts the peak strength in favor of the low-$\Erec$ tail.
The reduction in the high-$\Erec$ tail on the other hand, is due to the reduction of the cross section for larger scattering angles.

The low-$\Erec$ tail in the HF-CRPA result will affect results in terms of reconstructed energies in an analogous way as multi-nucleon emission processes.
This becomes clear from Fig.~\ref{fig:MMvsCRPA}, where the results obtained in the RPA, and RPA+np-nh of Martini~\etal.~\cite{Martini2012, Martini:recooscill}, are compared to the CRPA results.
Note that in the results of Martini~\etal.\ the reconstructed energy is defined with $E_B=0~\mathrm{MeV}$.
The compared $d(\E,\Erec)$ are hence not computed for the same $\Erec$ bins, but this mainly results in a shift of the distribution.

For $\E = 200~\mathrm{MeV}$, the np-nh strength does not contribute appreciably, apart from the shift the CRPA distribution is comparable with the RPA and RPA+np-nh results, although it is slightly broader.
The main difference between the RPA and CRPA distributions is most obvious for $\E=1~\mathrm{GeV}$.
The peak, and the high-$\Erec$ tail, are quenched in favor of the low-$\Erec$ tail. 
The np-nh distributions also feature heavy low-$\Erec$ tails due to the multinucleon knockout events, which also contribute in this kinematic region.
When comparing data in terms of $\Erec$ and $\E$ in the CRPA, we thus expect a redistribution of events towards lower values of $\Erec$ in a similar way as shown in Ref. \cite{Martini2012,Martini:recooscill}.

\subsection{MiniBooNE\label{sec:level23}}
We present the distribution of reconstructed energies in the CRPA model, and compare it to the results obtained in RPWIA and RFG approaches.
The success of the mean-field model in the low-energy regime is evident from comparison to inclusive electron scattering data \cite{RYCKEBUSCH1989,CRPAmod,Pandey:2016}.
The model is not fully relativistic however, it cannot be straightforwardly extended to arbitrarily large neutrino energies.
The goal of this work is therefore not to provide predictions of full reconstructed distributions, as this would also require a further assessment of two-nucleon knockout, resonances and other contributions to the $CC0\pi$ signal \cite{NuSTECWP,KatoriMartinireview}.
We will simply explore the $\Erec$-distributions for $\E < 2~\mathrm{GeV}$, to highlight the difference between the CRPA, RPWIA, and RFG results.

In this section, the number of events for a certain reconstructed energy $N(\Erec)$ given by Eq.~(\ref{eq:fluxd}), is computed with the MiniBooNE flux \cite{MBflux:2009}.
We will contrast this quantity with the flux-weighted total cross section $\Phi(\E)\sigma(\E)$, which we refer to as the number of events in terms of real energies $N(\E)$.
We thus compare the prediction in terms of energies, and contrast it with the predicted measurable distribution of reconstructed energies.
Note again that the results are divided by the total flux $\left(\int\,d\E\Phi(\E)\right)$ such that the distributions represent the flux folded cross section.

\begin{figure}[tb]
\begin{center}
\includegraphics[width=0.48\textwidth]{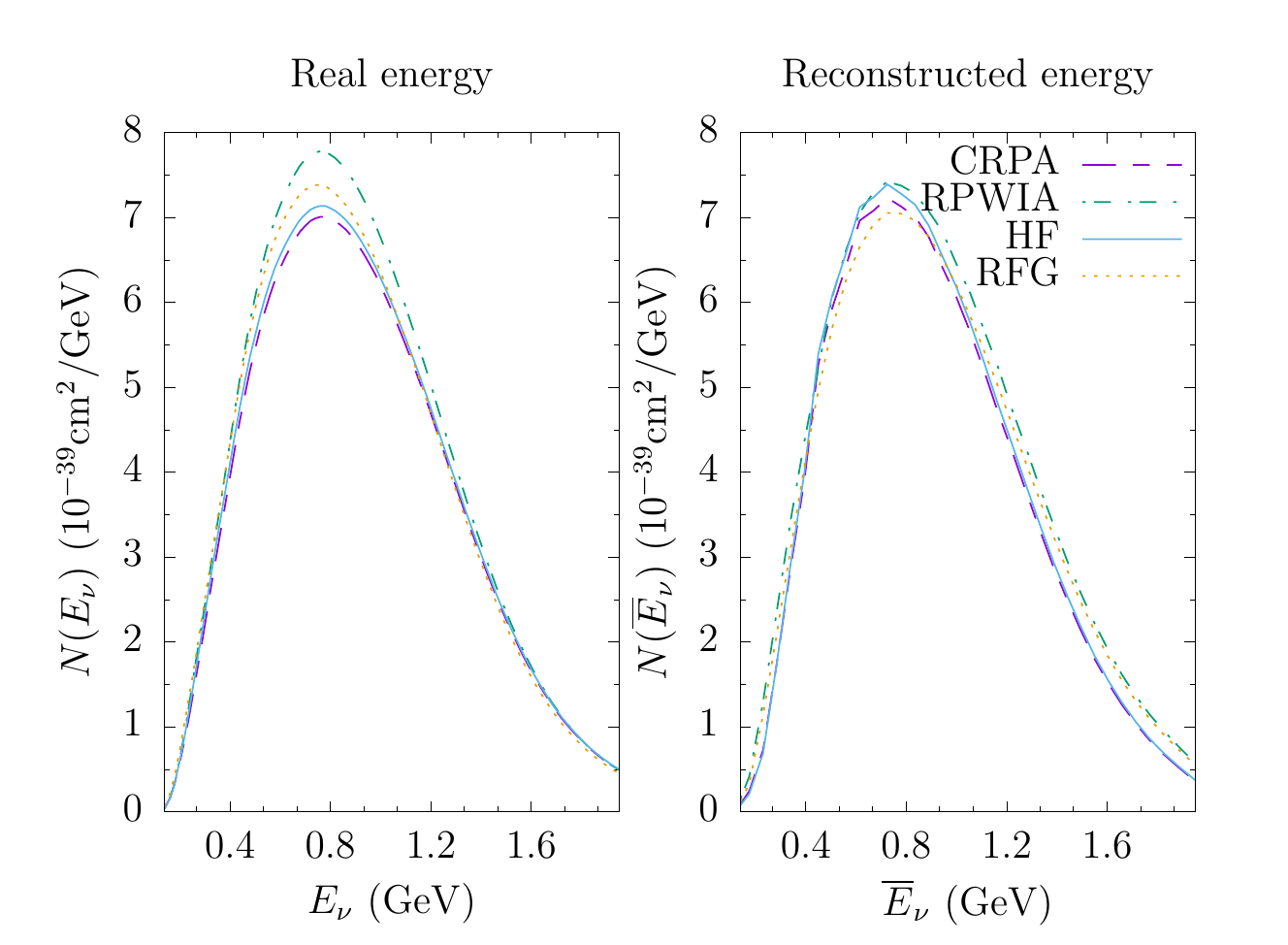}
\caption{Comparison of the total MiniBooNE flux folded cross section in terms of (reconstructed) energies.
The left panel shows the flux-weighted total cross section. The right panel shows the cross section in terms of reconstructed energies.}
\label{fig:CS_E_Erec}
\end{center}
\end{figure}
In Fig. \ref{fig:CS_E_Erec}, the flux-weighted total cross section is shown in the left panel.
The RFG model uses $k_F=221~\mathrm{MeV}$, and $E_s=34~\mathrm{MeV}$, similar to the values used in the MiniBooNE analysis \cite{MB:QECS}.
The RPWIA model predicts the largest cross section along the peak, this is due to the broad momentum distribution of the bound nucleons, and the lack of a proper treatment of the final state.
The final state plane wave is not orthogonal to the initial state wave function, which leads to unphysical contributions to the total strength for low energy transfers.  
The CRPA and HF model are comparable in strength, there is only a reduction owing to CRPA effects.

We see that the shape of the total integrated cross section $N(\E)$ is similar for all the models for this flux. 
The elastic distortion of the outgoing nucleon in the HF-CRPA models leads to a reduction in $N(\E)$ which is most obvious along the peak of the flux.
The RFG model can reproduce the position of the QE peak and overall strength of the cross section.
This shows that the specific shape of the double differential cross section is not of great importance after integration over the total leptonic phase space for fixed energies.
However $N(\Erec)$ depends on the lepton kinematics in a subtler way than $N(\E)$ as seen in Fig.~\ref{fig:cos}.
The spreading present in the double differential cross section leads to a different shape for $d(\E,\Erec)$, which affects the distribution of reconstructed energies $N(\Erec)$.

This is illustrated in the right panel of Fig.~\ref{fig:CS_E_Erec}, where $N(\Erec)$ is plotted.
We see that the models tend to agree in strength along the peak, even though this is not the case in the left panel.
These results show that the most interesting facet is that the HF-CRPA models are reshaped in a different way than the RFG and RPWIA models.
Indeed, when comparing $N(\E)$ to $N(\Erec)$ in the RFG and RPWIA, the peak is shifted slightly to the right, and is reduced in favor of the high-$\Erec$ tail.
The HF-CRPA models behave differently, $N(\Erec)$ is skewed towards lower reconstructed energies and the low-$\Erec$ peak is enhanced.

This is important in the analysis of neutrino-nucleus scattering experiments, where predictions of the cross section in terms of reconstructed energies are used to describe the signal and the background.
If $N(\Erec)$ obtained in an RFG can be made to agree to the CRPA results by tuning model parameters, this leads to a different interpretation in terms of real energies, because the reshaping, and therefore unfolding produces different shapes for $N(\E)$ for similar $N(\Erec)$ distributions.
In order to accurately describe the reconstructed data, models beyond an FG-based approach should be taken into account as inclusion of elastic FSI leads to a shape for $N(\Erec)$ which is not reproduced in the models using plane waves for the outgoing nucleon.
Even though $N(\Erec)$ obtained with different neutrino interaction models might agree for a certain energy distribution, this is not necessarily 
true for different fluxes. This will be illustrated in the next section.

\subsection{T2K\label{sec:level24}}
The T2K experiment is a long baseline neutrino experiment, using a neutrino beam originating from the J-PARC accelerator.
It uses two detectors, the near detector ND280 measures the events before oscillation, and the oscillated events are detected at the Super-Kamiokande (SK) detector $295~\mathrm{km}$ away. 
In the oscillation analysis, the experiment uses the cross section measured in the near detector to constrain the nuclear interaction model and the flux \cite{T2K:Measurement}.
The cross section is modeled using the RFG \cite{SmithandMoniz} implemented in the NEUT event generator \cite{Hayato:NEUT} and in addition to this, the 2p-2h contribution of Nieves~\etal.~\cite{NIEVES2013} is taken into account. 
Several parameters are fitted to the data measured in ND280.
Parameters pertaining to the RFG model include the axial mass $M_A$, the Fermi momenta for carbon and oxygen, and the binding energy for both nuclei \cite{T2K:Measurement}.
The systematic uncertainties on oscillation parameters measured in SK are determined by the constraints on these parameters which are fitted to the available near detector data.

In this section, we consider a single parameter, namely the binding energy of the RFG, implemented as an energy shift as previously explained.
A shift of $25~\mathrm{MeV}$ is commonly used to describe the cross section on carbon \cite{SmithandMoniz}, although the precise value and its interpretation depend on the specific implementation of the RFG model \cite{Bodek:Binding}.
 We vary this parameter and compare $N(\Erec)$ obtained in the RFG with the HF results, not taking into account CRPA corrections as it was seen in the previous sections that the influence on the reconstructed energy distributions is small. 
The cross section in terms of reconstructed energies is presented for both the near detector flux, and the predicted $\nu_{\mu}$ flux at SK, using a simple 2-neutrino oscillation prescription. 
Again, we limit the neutrino energy to values lower than $\E=2~\mathrm{GeV}$ in every calculation.
For simplicity, all calculations are done for a carbon nucleus.
The reconstructed energy is defined using Eq. (\ref{eq:erecbind}) with $E_B=25~\mathrm{MeV}$ fixed for every calculation.

\begin{figure}[tb]
\includegraphics[width=0.48\textwidth]{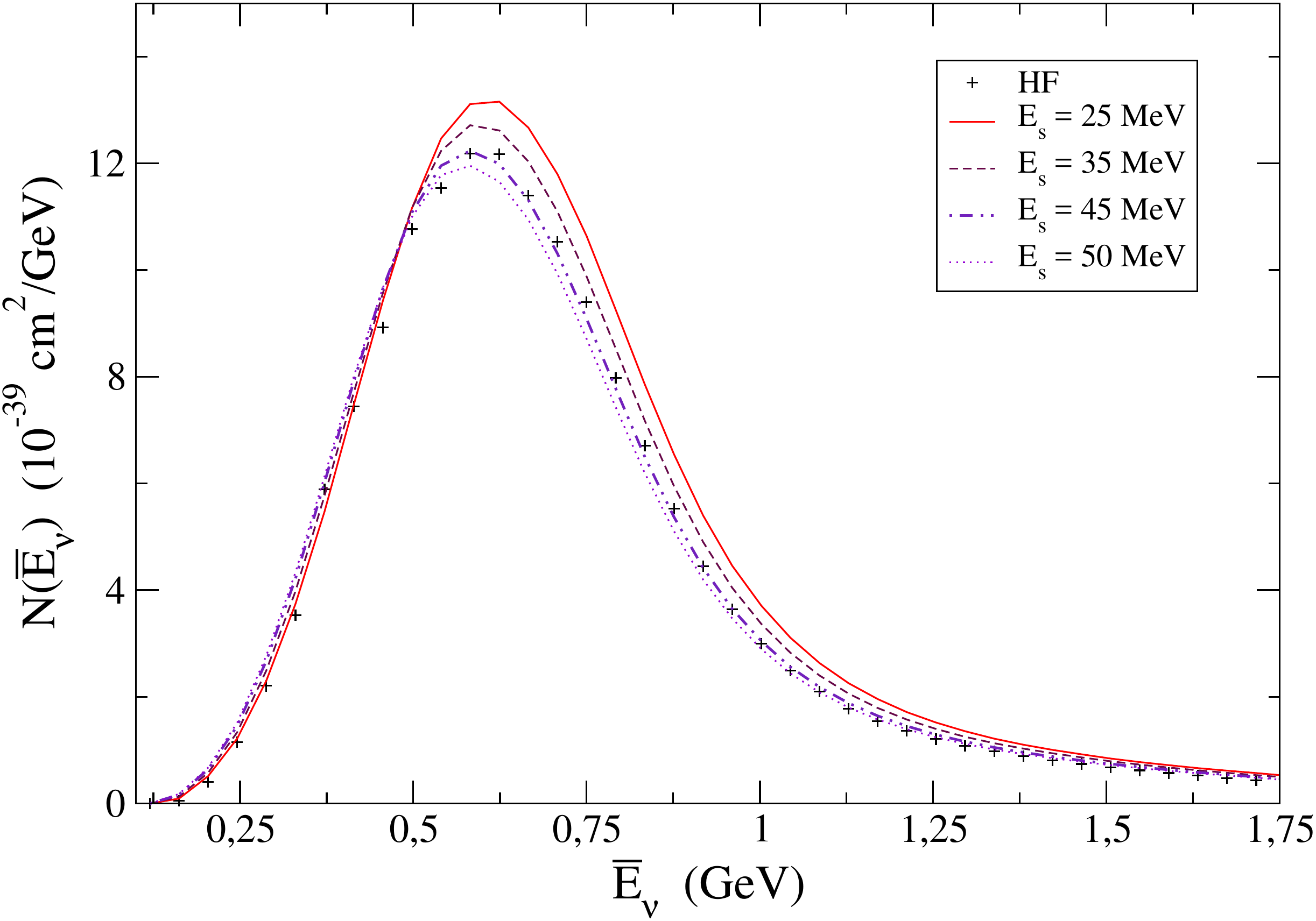}
\caption{Comparison of the cross section in terms of reconstructed energies. Weighted with the T2K ND flux.
The RFG calculations are labeled by their energy shift $E_s$.}
\label{fig:T2K_ND_Erec}
\end{figure}

When comparing the ND event distributions in terms of $\Erec$, shown in Fig. \ref{fig:T2K_ND_Erec} for the HF and for the RFG with different energy shifts, one sees that large values for the energy shift are required to reduce the peak strength of the cross section in order to be comparable to the HF result.
The best fit is obtained for $E_s \approx 45~\mathrm{MeV}$, in agreement with the value discussed in Ref.~\cite{Bodek:Binding}.
%For the ND, the minimal squared difference is obtained for a shift of $\approx 45~\mathrm{MeV}$

\begin{figure}[tb]
\includegraphics[width=0.48\textwidth]{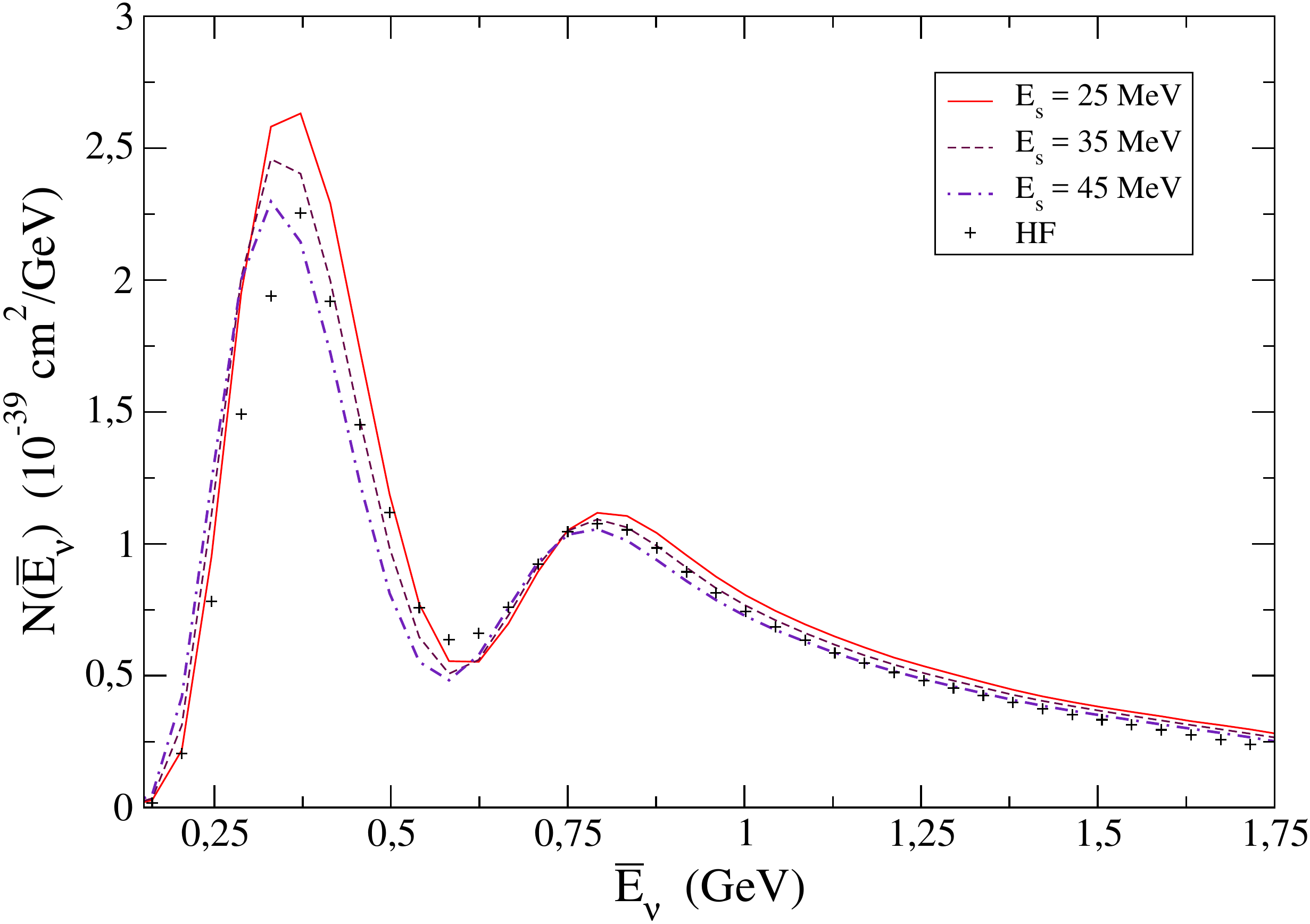}
\caption{Comparison of the cross section in terms of reconstructed energies, weighted with the T2K SK flux, with the two neutrino disappearance probability given by Eq. (\ref{eq:oscillated}).
Results are obtained in the HF approach, and with RFG models with different values of the energy shift $E_s$.}
\label{fig:T2K_SKos_Erec}
\end{figure}

The predicted cross section for an oscillated flux in the SK detector, calculated for carbon, is shown in Fig. \ref{fig:T2K_SKos_Erec}.
The disappearance probability of the muon neutrino was calculated using a 2 neutrino oscillation probability, with $\sin^2(2\theta_{23})=0.97$, and $\Delta m^2_{} = 2.32~10^{-3}\mathrm{eV}^2 $. The oscillated $\nu_{\mu}$-flux is then given by
\begin{equation}
\label{eq:oscillated}
\Phi(\E) =\left[ 1-\sin^2\left(2\theta_{23}\right)\sin^2\left(\frac{\Delta m^2L}{4\E}\right)\right]\Phi_{SK}(\E),
\end{equation}
where the predicted unoscillated flux at the far detector $\Phi_{SK}$ was taken from Ref.~\cite{T2K:flux2016}.
Compared to the ND distribution, where the peak strength and shape in the HF and RFG with a shift of $45~\mathrm{MeV}$ are in agreement,
we see a reduction of the cross section along the peak, but an increase of strength in the dip region and for the second peak. 
This can be explained again by considering the low-$\Erec$ tails for larger energies $\E$. The flux at larger energies contributes to the dip region through these tails.
The inclusion of np-nh has a comparable effect as shown in Ref. \cite{Martini:recooscill}.
The smaller cross section along the first peak can be mimicked by using a larger binding energy in the RFG.
This is however not true of the dip and the second peak, where a smaller binding energy $E_s$ is needed to match the HF prediction. 
This shows that even though both models may agree for a certain flux, due to different reshaping of the cross section these results are not necessarily suited to describe a different neutrino flux.
Changes of the energy shift in the RFG model have an effect on the position of the dip and peaks, which are important in determining the mass parameter $\Delta m^2$.

\section{Conclusion\label{sec:conclusions}}
A crucial problem in the oscillation analysis is the determination of the neutrino energy distribution in a detector, usually done through the reconstructed energy approximation.
An interaction model is necessary to describe the connection between the reconstructed and the genuine energy distributions.
In this work, we have investigated this connection considering QE one nucleon knock-out in the Hartree-Fock Continuum Random Phase Approximation, referred to as HF-CRPA, which incorporates in particular the mean field bound nucleons in the initial state and distortion of the outgoing nucleon wave function.
We have discussed the effect of the nuclear model by introducing the distribution $d(\E,\Erec)$ of reconstructed energies $\Erec$ around the genuine $\E$ value.
 In the RFG model the Fermi motion broadens this distribution in a  symmetric way around the real energy $E$.
 In the HF-CRPA model we have shown that the distribution of reconstructed energies around the real energy favors lower reconstructed energy values.
More specifically the distortion of the outgoing nucleon wave function leads to a pronounced low reconstructed energy tail, and the realistic treatment for low energy transfer shifts the peak of the distribution $d(\E,\Erec)$.
 We have investigated  the consequences for oscillation experiments, in particular for T2K.
 We aimed at reproducing the reconstructed energy distribution in the T2K near and far detectors predicted in the HF approach with an RFG.
 In the RFG the energy shift $E_s$ is taken as an adjustable parameter.
For the near detector flux a satisfactory fit is obtained for an energy shift of of $E_s=45~\mathrm{MeV}$.
 However for the oscillated distribution there is no value of the energy shift which, in the simple Fermi gas expression, fits our evaluation for the reconstructed energy distribution. 
Exporting parameters which reproduce near-detector data, to a different flux in the far detector, can in this way lead to a significant bias on oscillation parameters.
 The conclusion is that the sophistication of the Fermi gas model is not sufficient in this case.
Taking nuclear effects beyond the RFG into account could lead to a more satisfying understanding of the experimental data,
 it could for example remove a bias induced by the use of a constant binding energy in the experimental analyses.

 \begin{acknowledgments}
This  work  was  supported  by  the  Interuniversity   Attraction   Poles   Programme   initiated   by
the  Belgian  Science  Policy  Office  (BriX  network
P7/12)  and  the  Research  Foundation  Flanders
(FWO-Flanders), and by the Special Research  Fund,  Ghent  University.
The  computational resources (Stevin Supercomputer Infrastructure)  and  services  used  in  this  work  were  provided  by  Ghent  University,  the  Hercules  Foundation   and   the   Flemish   Government.
RGJ was partially supported by Comunidad de Madrid and UCM under the contract No. 2017-T2/TIC-5252.
M.M. acknowledges the support and A.N. the hospitality of the ``Espace de Structure et de r\'eactions Nucl\'eaires Th\'eorique'' (ESNT, \url{http://esnt.cea.fr} ) at CEA where part of this work was done.
\end{acknowledgments}
\bibliographystyle{apsrev4-1.bst}
\bibliography{Bibliography}% Produces the bibliography via BibTeX.

\end{document}